\documentclass[aip,apl,reprint,superscriptaddress,showpacs]{revtex4-1}

\usepackage{graphicx}
\usepackage{times}
\usepackage{amsmath}
\usepackage{subfigure}

\begin{document}

\bibliographystyle{aip.bst}

\title{Charge Imbalance and Crossed Andreev reflection in Py/Ta Devices}

\author{J.L.~Webb}
\email[email:~]{pyjlw@leeds.ac.uk}
\homepage[web:~]{http://www.stoner.leeds.ac.uk}
\affiliation{School of Physics \&\ Astronomy, University of Leeds,
Leeds, LS2 9JT UK}

\author{B.J.~Hickey}
\affiliation{School of Physics \&\ Astronomy, University of Leeds,
Leeds, LS2 9JT UK}

\author{G.~Burnell}
\affiliation{School of Physics \&\ Astronomy, University of Leeds,
Leeds, LS2 9JT UK}

\begin{abstract}
Crossed Andreev reflection in a lateral spin valve geometry device is an aspect of considerable recent interest, particularly with regards to Cooper pair splitting experiments to realize solid state quantum entanglement. In this work, devices are fabricated consisting of two ferromagnetic permalloy (Py) electrodes contacted by a overlaid perpendicular superconducting Ta electrode, of variable lateral in-plane offset within the expected BCS superconducting coherence length $\xi_0$. Experimental electrical transport measurements are presented of local and nonlocal conduction, showing characteristics of non-local or crossed Andreev reflection in the superconducting state at low temperatures in addition to nonlocal effects divergent for T$\rightarrow$T$_c$ attributed to nonlocal charge imbalance. 
\end{abstract}

\date{\today}
\maketitle
Andreev reflection \cite{Andreev1} (AR) is a charge transfer process through a single superconductor/normal state metal (S/N) interface at energies less than the superconducting energy gap $\Delta$. Conduction through the interface is achieved through retroreflection of - in the case of an incident electron - a hole with opposite wavevector $\textbf{k}$ into the normal metal and opposite spin to the incident electron. AR is thus both spin dependent and dependent on the nature of the interface. Retroreflection occurs 
on the range of the BCS superconducting coherence length $\xi_0$ several hundred nanometers in materials such as Al. Due to the large length scale involved, it is possible for the retroreflection to occur at a second S/N interface at an adjacent (detector) electrode than that (the injector) at which the initial incidence occurred. This is possible if the adjacent electrode is fabricated within $\xi_0$, and is termed non-local or crossed Andreev reflection (CAR). If ferromagnetic (FM) electrodes are used, this nonlocal process is dependent on the magnetization state, either parallel(P) or antiparallel(AP), of the respective electrodes \cite{2004PhRvL..93s7003B}. CAR presents a possibility of to achieve solid state quantum entanglement by Cooper pair splitting, of current experimental and theoretical interest \cite{2009NatPh...5..393C}.

As yet, such processes have only been observed clearly in a limited range of superconductors such as Al and In \cite{2009PhRvB..80v0512A} with a long $\xi_0$ and primarily at temperatures in the low hundreds of mK. This is due to the difficulty in isolating the CAR process from competing processes, such as elastic co-tunneling (EC) from injector to detector via an intermediate state in the superconductor or nonlocal charge imbalance (CI) effects arising from quasiparticle injection. It is desirable to extend work toward superconductors with shorter coherence lengths, and at higher temperatures, to further understand the processes, to more easily generate them and and to combine such devices with existing superconducting devices, based on higher T$_c$ materials or similar spintronic devices such as the nonlocal spin valve (NLSV) devices. 

\begin{figure}
\includegraphics[width=1.0\linewidth]{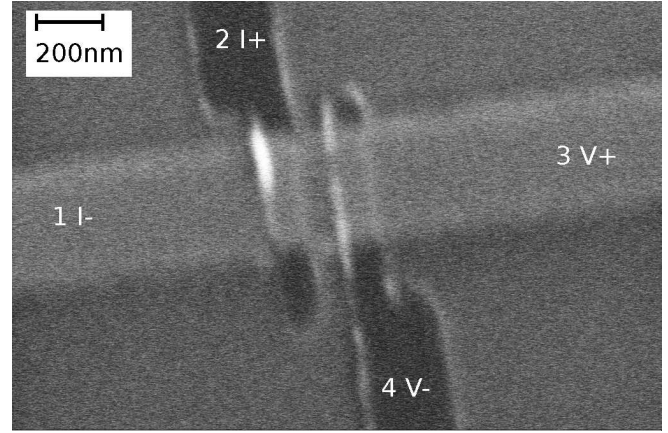}\\
\caption{SEM image of example device, with two Py electrodes (left the injector, right the detector). Electrode widths were designed (actual) 100nm (93nm and 96nm). A 300nm width Ta electrode is patterned across both electrodes. }
\label{fig1}
\end{figure}

In this paper, we present measurements of non-local conduction through superconducting tantalum, in a conventional NLSV geometry consisting of two ferromagnetic electrodes in contact with a perpendicular superconducting strip as shown in Fig. \ref{fig1} with variable lateral separation. A range of devices were fabricated by electron beam lithography (Raith 50 system), consisting of two 100nm width Py electrodes with separations ranging between 150nm to 75nm in contact with a 300nm width Ta track. Py deposition was performed by DC magnetron sputtering in a system with base pressure of 3x10$^{-9}$torr at 36W in 2.5mTorr of Ar gas. Further fabrication of the device was performed ex-situ by a second EBL mix-and-match overlay step. This necessitated cleaning of the Py/Ta interface with 40s of Ar$^{+}$ milling at 410-450V DC followed by rapid in-situ Ta growth at 0.42nm/s at 50W DC/2.5mTorr Ar, performed in a second machine with system base pressure 3.2x10$^{-8}$torr. The Ar$^{+}$ stage was found necessary to remove residual resist and oxide formation that would otherwise act to inhibit both efficient electrical conduction and destroy any spin dependent process sought.

Measurement was performed in a continuous flow He-4 cryostat cooled to a minimum of 1.4K, and in an adiabatic demagnetization refrigerator down to a minimum of 200mK, with current injection (and local voltage detection) from 1-2 (Fig. \ref{fig1}) and non-local voltage detection 3-4. Measurement was performed using an applied DC current $I_{dc}<$200$\mu$A, with voltage measurement simultaneously locally and non-locally. In addition, a standard lock-in amplifier technique using an AC current $I_{ac}$=0.25$\mu$A at 9.99Hz with DC offset $I_{dc}$ was used to measure differential resistance $dR_{nl}$. A low frequency was selected in order to reduce the voltage induced nonlocally by inductive coupling between the two magnetic elements. Here we present data from 2 devices (1,2) from 2 device sets (A,B) grown separately by the above method.
 
\begin{figure}
\includegraphics[width=1.0\linewidth]{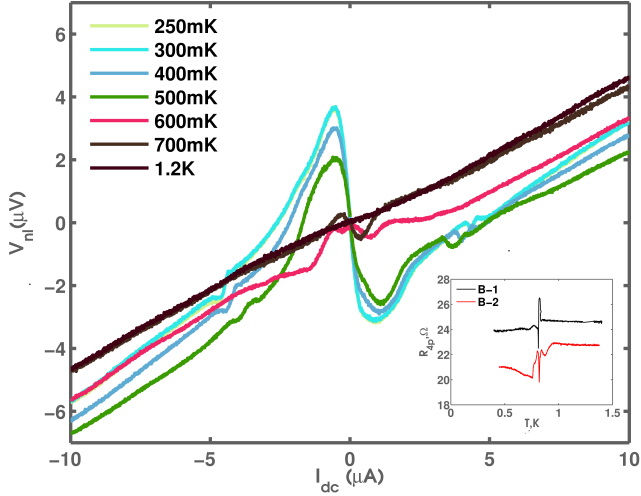}\\
\caption{DC $V_{nl}$ vs. $I_{dc}$ at T=250mK-1.2K. A linear fit to the data yielded a finite nonlocal background resistance of $R_{bg}$=0.45$\Omega$. This was be attributed to diffusive current flow into the detector electrode, due to the presence of a normal state region within the device. When subtracted, a negative nonlocal voltage consistent with CAR of $V_{nl}$=3.5-3.7$\mu$V was observed in the superconducting state. (Inset) 4-probe resistance measurement R$_{4p}$ of Ta electrode bridge inbetween the magnetic electrodes, indicating superconducting transition and presence of a nonzero resistance below T$_c$ }
\label{fig2}
\end{figure}

Fig. \ref{fig2} shows the DC nonlocal voltage $V_{nl}$ at the detector electrode (3-4) as a function of $I_{dc}$ through the injector for device A-1 for a range of temperatures between 250mK (superconducting state) and 1.2K (normal state). As T was reduced, a perturbation from linear behavior in the normal state and at high $I_{dc}$ emerged for energies corresponding to below the gap, estimated at the edge of the perturbation region at I$_{dc}\approx$2$\mu$A at $\Delta\approx$0.12meV. The background nonlocal resistance $R_{bg}=V_{nl}/I_{dc}\approx$0.45$\Omega$ was initially attributed to nonlocal charge imbalance. However, the effect was temperature-invariant and extant in the normal state, suggesting instead diffusive conduction through a normal state region between the FM electrodes, indicated by a finite resistance in the Ta electrode between the ferromagnetic electrodes (inset, Fig. \ref{fig2}). On performing a linear fit at 1.2K and subtracting from all data, the resulting perturbation represented a negative nonlocal voltage at the detector electrode characteristic of CAR \cite{2006AIPC..850..875B} at a peak of 3$\mu$V at 250mK. The independence of R$_{bg}$ to temperature and increasing injector current suggested the formation of a normal state region from magnetic stray field effects rather than originating from thermal suppression of the superconductivity. 

\begin{figure}
\includegraphics[width=1.0\linewidth]{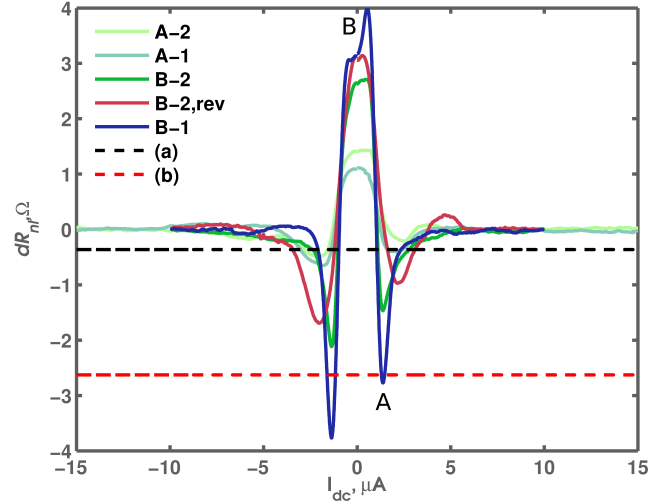}\\
\caption{Nonlocal differential resistance ($dR_{nl}=dV_{nl}/dI$) measured by lock-in using $I_{ac}$=0.25$\mu$A as a function of applied DC offset $I_{dc}$ for all devices at T=600mK, exhibiting negative minima (label A) and zero I$_{dc}$ peak (B). Current injection was through 1-2 and voltage detection 3-4 (\textit{rev} denotes the reverse case). An Ohmic background $dR_{bg}$ was subtracted from from the data, of magnitude indicated by dashed lines (a) and (b) at -0.36$\Omega$ and -2.62$\Omega$.}
\label{fig3}
\end{figure}

In addition, measurement of the differential resistance $dR_{nl}$ was performed at $I_{ac}$=0.25$\mu$A combined with a variable DC offset current $I_{dc}$. The result for all devices can be seen in Fig. \ref{fig3}. Again, the background effect attributed to diffusive current flow was subtracted. The perturbation effects observed were larger in device Set B than A indicative of the effect originating from the configuration of the device as opposed to bulk material properties. The data suggested two possible explanations for the features shown. Firstly, that the negative minima and zero $I_{dc}$ (zero bias) peak arose from competition between CAR and EC, as observed by Russo et al. \cite{2005PhRvL..95b7002R}, Beckmann et al. \cite{beck1} and Kleine et al. \cite{2009EL.....8727011K} and calculated theoretically by Yeyati et al. \cite{2007NatPh...3..455Y}. Secondly, that the effect was a product exclusively of nonlocal CI, not CAR. Negative $dR_{nl}$ at finite applied current was observed in previous work on nonlocal CI by Cadden-Zimansky et al. \cite{2007NJPh....9..116C}\cite{2006PhRvL..97w7003C}. However, in this work the CI-attributed $dR_{nl}$ reached a finite bias peak of height greater than that at zero injector bias followed by a negative minima many times larger than the preceding peak. We do not observe this and instead, we observe features consistent with the first explanation. 

\begin{figure}
\includegraphics[width=1.0\linewidth]{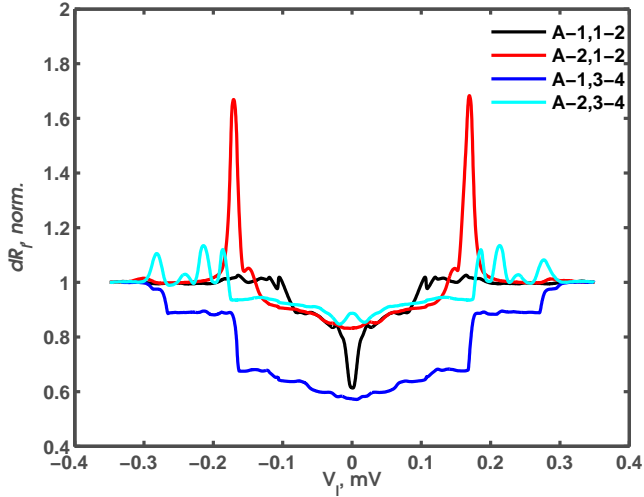}\\
\caption{Local differential resistance $dR_{l}$=dV$_{l}$/dI  as a function of local bias V$_l$ through both injector and detector electrodes of the devices in set A. The Ta was superconducing, exhibiting a minima in resistance observed within the expected bulk $\Delta$ attributable to Andreev reflection. The varied features observed - such as the peaks in $dR_{l}$ at $\approx$5$\mu$A - were attributed to localized spatial variations in $\Delta$ at each of the injector/detector interfaces and due to CI effects at higher $I_{dc}$}
\label{fig4}
\end{figure}

As can be seen in Fig. \ref{fig3} a considerable asymmetry was observed between both device sets and between injector/detector choice. This may be understood from analysis of the local differential resistance $dR_l=dV_l/dI$ measurements exemplified for the devices in set B in Fig. \ref{fig4}. Changes in $dR_l$ were attributed to CI and AR at the superconductor-ferromagnet interface and thus the differences and range of features within each most likely represented either local variation of $\Delta$ due to the extended non-point contact nature of the interface, or from single junction CI effects as characterized by Chein et al. \cite{PhysRevB.60.3655}. Such features rendered the data not fitable using the standard BTK \cite{1982PhRvB..25.4515B} approach suited only to a point contact. Local variation in $\Delta$ was expected to arise from the inverse proximity effect and stray field from the FM electrodes. Such a situation, with local suppression of the superconductivity, support the hypothesis above as to the origin of $R_{bg}$ but would not necessarily impede other effects if the Ta bridge adjoining the electrodes was sufficiently well superconducting (i.e.. with local $\Delta\rightarrow\Delta_{bulk}$). 
 
\begin{figure}
\centering
\subfigure[]{
\includegraphics[width=1.0\linewidth]{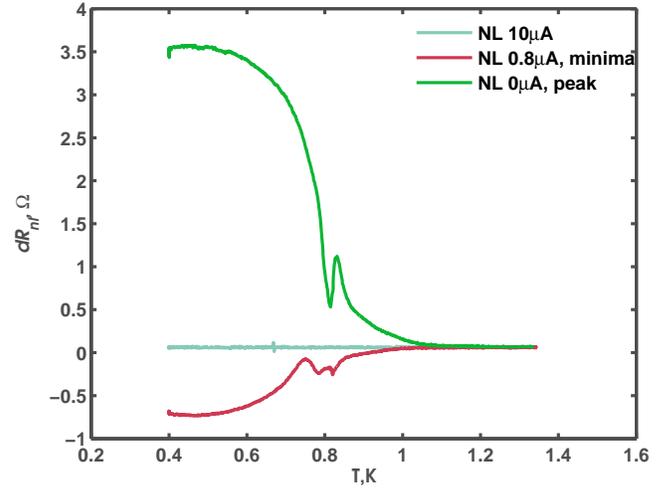}\
\label{fig:fig5a}
}
\subfigure[]{
\includegraphics[width=1.0\linewidth]{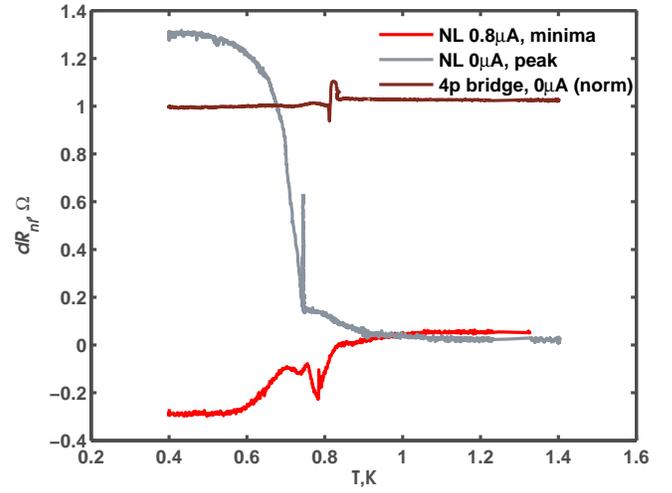}\
\label{fig:fig5b}
}
\caption[]{Nonlocal effect temperature dependence: a) $dR_{nl}$ for device B-1, b) $dR_{nl}$ for device B-2 as a function of temperature T, from 400mK to 1.4K. In a) the negligible effect at high I$_{dc}$=10$\mu$A is shown. In b) the normalized 4-probe (4p) resistance of the Ta bridge region between the magnetic electrodes is overlaid to indicate the position of T$_c$, described by the step in resistance at $\approx$800mK }
\end{figure}

In order to separate EC/CAR effects from CI, the $I_{dc}$=0 peak and negative minima points of Fig \ref{fig3}. were investigated as a function of temperature. This can be seen in Figs. \ref{fig:fig5a} and \ref{fig:fig5b} for device set B in AC measurement mode. At low temperatures ($<$600mK), the nonlocal measurements tended to a temperature independent state, diverging away from this as T$\rightarrow$T$_c$ at around 800mK. The range over which charge imbalance effects occur, the charge imbalance length given by \cite{2007NJPh....9..116C}\cite{springerlink:10.1007/BF00115264} $\Lambda_Q = \sqrt{D\tau_Q}$ where $\tau_Q \propto{1/\Delta(T)}$ is also divergent as $\Delta(T)\rightarrow0$. The temperature independent state for T$\rightarrow$0 likely represents the state $\Lambda_Q<<$L, the inter electrode separation. It may thus be deduced that CI effects were minimized at or below the 600mK temperature which prior transport measurements in Figs. \ref{fig2}-\ref{fig4} were performed at. The magnitude of $dR_{nl}$ exhibited a decay with T towards the normal state as opposed to a divergent increase in $dR_{nl}$ for T$\rightarrow$T$_c$ as observed by Beckmann et al. \cite{2004PhRvL..93s7003B} and others. This difference may be attributed to the existence of other effects (i.e. - EC or CAR) to produce a finite $dR_{nl}$ at low temperature overlaid onto the typical CI divergent behavior observed in literature, manifested as the positive peak in nonlocal resistance close to T$_c$. A similar divergent effect was observed for Al devices also fabricated, which are reported elsewhere. 

A 4-probe measurement (current application 1-3, voltage detection 2-4 on Fig. \ref{fig1}) was performed on the Ta electrode joining the two ferromagnetic electrodes. This yielded a finite resistance at low temperature when a zero-resistance superconducting state was expected, consistent with the existence of a region for which $\Delta\rightarrow$0 to produce $R_{bg}$ at T$<$T$_c$. However, it was clear a superconducting region existed between the two electrodes sufficient to produce nonlocal superconducting effects, its presence indicated by a transition to a lower resistance state at $\approx$800mK. The precise localized extent of such a normal state region could not be directly probed, but could be inferred from transport measurements as being adjacent to the Py electrodes, arising through suppression of local superconductivity by stray magnetic field. 

\begin{figure}
\includegraphics[width=1.0\linewidth]{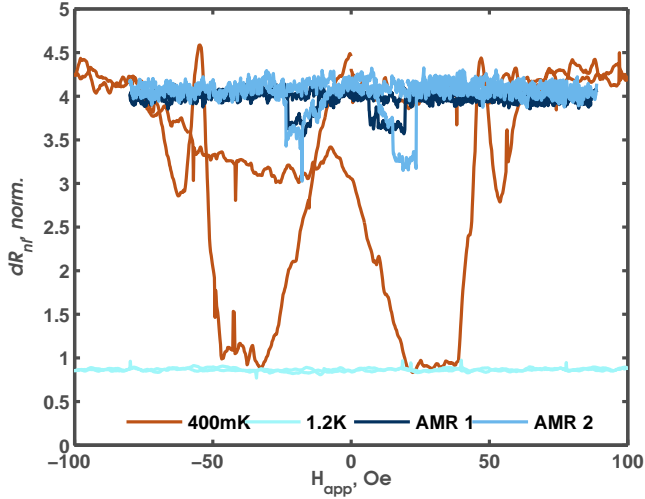}\\
\caption{Magnetic field dependence of $dR_{nl}$ at $I_{dc}$=0 and 400mK/1.2K in the superconducting/normal state. The data is presented normalized to the 1.2K normal state value (=1). Overlaid are scaled AMR measurements taken I$_{dc}$=10$\mu$A, indicating the magnetic switching of each electrode. Switching to the 1.2K normal state value was coincident with electrode coercive field. }
\label{fig6}
\end{figure}

In order to separate CAR and EC, a magnetic field H$_{app}$ was applied to the device along the length of the electrodes to induce the required AP/P magnetization states. This data may be seen in Fig. \ref{fig6} is terms of $dR_{nl}$ normalized to the differential resistance in the normal state. For I$_{dc}$ = 0 , the peak $dR_{nl}$ fell as H$_{app}$ was swept. Such a change would be anticipated on switching between the AP/P magnetic states. However, on performing a DC AMR measurement through each electrode at I$_{dc}$ = 10$\mu$A, the coercive fields (indicated by the AMR minima) coincided only with the low field fall in $dR_{nl}$. Further, the entire electrodes did not form a full AP state as indicated by considerable overlap in their AMR data. In addition, the minima of $dR_{nl}$ coincided with the value taken at 1.2K in the normal state, as shown on Fig. \ref{fig6}. This suggested the change in $dR_{nl}$ could only be attributed to local suppression of the superconductivity by stray magnetic field. That the return to the superconducting state did not occur until higher field was attributed to a region of higher coercive field in the magnetic electrode adjacent to the Ta, arising from localized defect pinning, or potentially through changes in current and heat distribution in the device after the switch to the normal state. The return did not occur however on a timescale consistent with device thermal dissipation. 
That the superconducting state could be modified in this manner - and does not immediately return to the same superconducting state, a feature demonstrated for H$_{app}<$0 in Fig. \ref{fig6} - raises an important point of caution for the accurate differentiation of EC/CAR using magnetization states, especially in the case where the electrode coercive field required to create such states can exceed the critical field of the superconducting strip.
	
In conclusion, devices have been fabricated in a conventional Ta/Py lateral spin valve geometry and nonlocal effects measured, including negative nonlocal voltage of 5-6$\mu$V a zero bias non-local conductance peak and finite bias negative minima at energy scales characteristic of EC/CAR. This was realized by subtraction of a background resistance R$_{bg}$ attributed to local suppression of superconductivity in the Ta electrode that permitted nonlocal current diffusion, which is attributed here to the stray field and proximity effect of the magnetic electrodes adjacent to the Ta due to the temperature and current independence of R$_{bg}$ observed, contrary to what would be seen for purely thermal suppression. This local suppression of the superconductivity was insufficient to inhibit the nonlocal superconducting effects. Temperature invariance of $dR_{nl}$ at T$\rightarrow$0 was indicative of the suppression of CI, which emerged in the form of a divergent effect as T$\rightarrow$T$_c$. These results contribute further experimental data and understanding on nonlocal charge imbalance and nonequilibrium states in a lateral spin valve geometry. A note of caution is raised through measurements with an applied magnetic field with respect to the influence of field induced local suppression and its persistence, with respect to differentiating between EC/CAR in the P and AP magnetic states.	

This work was supported by a UK Engineering and Physical Sciences Research Council Advanced Research Fellowship EP/D072158/1.
\bibliography{all3.bib}

\end{document}